\documentstyle[prl,aps,tighten,multicol,epsf]{revtex}
\begin{document}

\title{Long--range interacting rotators:
connection with the mean--field approximation} 

\author{Francisco Tamarit$^1$
\thanks{Member of the National Research Council, CONICET(Argentina)}
 and Celia Anteneodo$^2$\thanks{Mailing author} }
\address{$^1$
         Facultad de Matem\'atica, Astronom\'\i a y F\'\i sica,
         Universidad Nacional de C\'ordoba, \\
         Ciudad Universitaria, 5000 C\'ordoba, Argentina \\
         {\rm e-mail: tamarit@fis.uncor.edu } \\
	 $^2$
	 Instituto de Biof\'{\i}sica, Universidade Federal do Rio de
         Janeiro, \\ 
         Cidade Universit\'aria, CCS Bloco G, CEP 21941-900,
         Rio de Janeiro, Brazil \\
         {\rm e-mail: celia@cbpf.br} }
\maketitle
\begin{abstract}
We analyze the equilibrium properties of a chain of ferromagnetically 
coupled rotators which interact through a force that decays as 
$r^{-\alpha}$ where $r$ is the interparticle distance and $\alpha\ge 0$. 
Our model contains as particular cases the mean field limit ($\alpha=0$) 
and the first--neighbor model ($\alpha \to \infty$). 
By integrating the equations of motion we obtain the microcanonical 
time averages of both the magnetization and the kinetic energy. 
Concerning the long--range order, we detect three 
different regimes at low energies, depending on whether $\alpha$
belongs to the
intervals $[0,1)$, $(1,2)$ or $(2,\infty)$. 
Moreover, for $0 \le \alpha < 1$, the microcanonical averages agree,
after a simple scaling, with those  obtained in the canonical ensemble
for the mean--field XY model.
This correspondence offers a  mathematically tractable and computationally
economic way of dealing with systems governed by slowly decaying long--range
interactions.
\end{abstract}

\begin{multicols}{2}

\narrowtext

One of most important questions in statistical mechanics refers
to the connection between dynamics and thermodynamics:
To what extent a suitable ensemble average allows to predict the time
average of a physical observable performed by our instruments in
the laboratory? Or, in other words, which are the mechanical specifications 
of those systems to which the results of statistical physics can 
be applied? Within this context, while ergodicity and mixing have been
analyzed intensively  in the literature, there is another important point
which has not deserved the same degree of attention, that is the
possibility of defining a thermodynamically suitable energy function.
In fact, for systems governed by sufficiently long--range 
interactions decaying as $r^{-\alpha}$ with the interparticle distance $r$,
there results a {\em non--extensive} Hamiltonian, i.e., 
the energy per particle diverges in the thermodynamics limit 
$N \to\infty$ \cite{Thirring,Tsallis}.  
Gravitational ($\alpha=1$, $d=3$) and monopole-dipole
($\alpha=2$, $d=3$) interactions are only two well known instances
among many others.
Furthermore, such forces are particularly interesting since they 
can lead to equilibrium behaviors different 
from those observed in short--range systems and even give place to 
phase transitions otherwise absent, even in the $d=1$ case. 

Our aim here is to investigate how to deal with systems governed by 
long--range interactions by analyzing a simple but rich prototype with
adjustable $\alpha$.
The main goal of this letter is to show that the mean-field 
limit ($\alpha = 0$) is able to describe the thermodynamics
in the whole range $0<\alpha<1$.
The model consists in a  one dimensional chain of $N$ 
interacting rotators with periodic boundary conditions. Each  
rotator moves on the unit circle and therefore it is fully described by 
the angle $-\pi < \theta_i \le \pi$  and 
its conjugate momentum $p_i$ (with $i=1, \ldots,N$). 
The dynamics of the chain is governed by the following Hamiltonian
\begin{equation}  \label{hamiltonian}
H = \frac{1}{2} \sum_{i}^{N} p_{i}^{2} \, +  \frac{1}{2} \sum_{i\neq j}
\frac{1 - \cos(\theta_{i} - \theta_{j} )}{r_{ij}^{\alpha} }
  \; \equiv \; K + U ,
\end{equation}
where, without loss of generality, we have chosen unitary moments of
inertia for all the particles. 
Here $r_{ij}$ measures the minimal 
distance between rotators $i$ and $j$ along the chain.
The Hamiltonian (\ref{hamiltonian}) describes a classical 
inertial XY  ferromagnet. 
$K$ and $U$ denote the kinetic and potential energies respectively. 
The equations of motion ruling this dynamical system are:
\begin{eqnarray}
\dot{\theta_{i}} & =& p_{i} \\
\dot{p_i} & = & -\sum_{j\neq i}
\sin{(\theta_i - \theta_j)} /r_{ij}^{\alpha}\, .
\end{eqnarray}

We associate to each particle  a spin vector
\begin{equation}
{\bf m}_{i} = ( \cos{\theta_{i}},\sin{\theta_{i}})
\end{equation}
and define the total magnetization of the system as:
\begin{equation}
{\bf M} = \frac{1}{N} \sum_i {\bf m}_{i} \, .
\end{equation}
The long--time behavior of $\bf{M}$ determines whether the system orders 
(${\bf M}\neq 0$) or not (${\bf M}=0$). 

We also introduce a time dependent temperature $T(t)$ as 
$T(t) = (2/N)<K>(t)$, where $<\ldots >$ denotes a time
average performed over the time interval $(0,t)$. By calculating the
long--time behavior of $T$ as a function of the total energy of the
system one gets the {\em caloric curve} $T$ vs. $E/N$ from which
the specific heat function is extracted.

Note that the $\alpha \to \infty$ limit yields the  
first--neighbor case while $\alpha = 0$ represents the mean--field version. 
The later case has a correspondence with the 
model known in the literature as hamiltonian mean--field XY (HMF), 
provided the potential energy (thus the strength of the interactions) 
is scaled by the number of particles $N$, i.e., in the HMF model 
$U=\frac{1}{2N}\sum_{i\neq j}[1-\cos(\theta_{i}-\theta_{j})]$. 
The HMF model has received special attention during the last years and very
interesting results have been obtained both for its equilibrium and
non--equilibrium properties\cite{Antoni,Vito1,Vito2}.
This model can be solved analytically within the canonical ensemble 
formalism\cite{Antoni}. It is found that there is a critical specific energy at 
which the system suffers a phase transition separating a paramagnetic 
high energy phase (with ${\bf M}=0$) from a ferromagnetic low energy one 
(with ${\bf M}\neq 0$). 
More precisely, $E/N=T/2 + (1-{\bf M}^2)/2$, with $|{\bf M}|=yT$ 
where $y$ is the value maximizing $-y^2T/2+\ln{I_0(y)}$, being $I_0$  the  
modified Bessel function of order $0$. 
For $E/N>3/4$ ($T>1/2$) the only solution is $y=0$ while below 
the critical energy the null solution becomes unstable and 
a new stable solution appears giving place to a second order phase transition. 
On the opposite limit of first--neighbor interactions ($\alpha \to \infty$), 
when $U=\sum_{i}[1 - \cos(\theta_{i+1} - \theta_{i})]$, 
one does not observe an order--disorder thermodynamical 
transition\cite{Escande}. 
In this case, $E/N=T/2+1-I_1(1/T)/I_0(1/T)$, where $I_n$ stands for 
the modified Bessel function of order $n$.  
The dynamics generated by the Hamiltonian (\ref{hamiltonian})
has been analyzed before\cite{Celia}. 
In the limit $N \to \infty$ and for energies above
a critical value, the maximal Lyapunov exponent was shown
to vanish for $0\le \alpha < 1$, while it tends to a finite
value otherwise.

For arbitrary values of $\alpha$ we have  integrated numerically the 
set of equations of motion  (2) and (3) using a fourth order symplectic
method\cite{Yoshida} with 
a fixed time step selected so as to keep the energy 
constant within an error $\Delta E / E$ of order $10^{-4}$. Initial 
configurations ($t=0$) were chosen as follows: all the angles were set 
to zero and the momenta were chosen at random from the uniform 
distribution with zero mean value. Next, all the momenta were scaled in
order to attain the desired total energy $E$. 
We varied the size of the systems from $N=100$ to $1600$ in order to analyze 
finite size effects \cite{algorithm}.  
Once elapsed a transient (that depends both on the size of the system and 
on the total energy $E$), we computed the time averages of both the 
magnetization and the kinetic energy looking for their asymptotic behavior.

We start considering the region $0 \le \alpha < 1$. 
Systems characterized by values of $\alpha$ within that interval 
do not have a well defined Hamiltonian in the sense that,
in the thermodynamics limit $N\to\infty$, the potential energy diverges.
One of the goals of this work is precisely to establish a connection 
between our microcanonical temporal averages obtained for $\alpha$ inside 
that range and those obtained theoretically by averaging in the 
canonical ensemble the HMF model\cite{Antoni}. 
 
Fig. 1a displays the  modulus of the magnetization $|{\bf M}|$ as a 
function of the energy per particle $E/N$ for $\alpha = 0.5$ and
different system sizes. We observe that the system undergoes
a phase transition from a ferromagnetic low energy 
phase with ${\bf M}\neq 0$ to a paramagnetic high energy phase 
where ${\bf M} = 0$. As expected, the critical specific energy diverges 
as $N$ increases, due to the lack of {\em extensivity}
of the system. In Fig. 1b we present the same results
but now with the specific energy scaled by a factor $\tilde{N}$ that
depends both on the system size $N$ and on $\alpha$. 
To understand the origin of this scaling, let us stress that
the lack of extensivity in the range $0 \le \alpha \le 1$ emerges
as a consequence of the divergence of the specific potential energy upper bound
in (\ref{hamiltonian}). $\tilde{N}$ is nothing else than 
twice the value of this upper bound:
\begin{equation} \label{Ntilde}
\tilde{N} \, \equiv \, 2 \sum_{r=1}^{N/2}\frac{1}{r^{\alpha}} 
\, \approx \,
2 \int_{1/2}^{N/2} dr \,r^{-\alpha} \, = \,
2^{\alpha} \; \frac{N^{1-\alpha} -1 }{1-\alpha}\; ,  
\end{equation}
For $N\to \infty$, $\tilde{N}$ behaves as
\begin{equation}
\tilde{N}(\alpha)\sim \Biggl\{
\begin{array}{ll}
2^{\alpha} \; \frac{1}{1-\alpha} N^{1-\alpha}& \mbox{for}   \qquad 0\le\alpha< 1 \\
2 \; \ln{N}                                  & \mbox{for}   \qquad \alpha = 1  \\ 
\Theta(\alpha) \frac{1}{\alpha - 1}          & \mbox{for}   \qquad \alpha > 1  \\
\end{array}
\Biggr.  \label{asym}
\end{equation}

\noindent where $\Theta(\alpha)$ is a function of $\alpha$ which, for
$\alpha\to\infty$ goes to $2(\alpha-1)$. Note that all the 
curves for different $N$ collapse into a unique one, despite small
discrepancies around the critical value.
Similar $\tilde{N}$-scaling collapse was already observed for
magnetic systems\cite{Sergio,Sampaio} 
and also for systems governed by interactions of 
the Lennard-Jones type\cite{Jund}.
In all these cases, 
although the $\alpha$--dependent prefactor in (\ref{asym}) is model
dependent, the behavior of $\tilde{N}$ with $N$ is invariant.  

Fig. 1c plots $T/\tilde{N}$ vs. the scaled specific energy  
$E/(N \tilde{N})$ (caloric curve)  also for $\alpha = 0.5$ 
and the  same system sizes considered in Figs. 1a and 1b. 
Here again the $\tilde{N}$-scaling leads to data collapse.
It is worth here to stress that around the critical energy this plot 
depends strongly on the equilibration transient, the size of the
system and the initial configuration adopted. 
Our results within this parameter range seem to indicate the existence 
of a first order phase transition, with the high energy phase coexisting 
with the ordered one. 
An analogous behavior, already reported for the HMF 
model\cite{Vito1}, where the transition is second order, 
was believed  to be a purely microcanonical result 
reflecting the existence of long living quasistationary nonequilibrium
states whose lifetimes increase with $N$. This seems to be also the 
case for any $0\le \alpha <1$ since the discrepancy around the 
transition for fixed energy and fixed size is attenuated by averaging over 
larger time intervals. 

Fig. 2 exhibits $|{\bf M}|$ vs. $E/(N\tilde{N})$ 
for  $ \alpha =0.25$, $0.5$ and $0.75$, all for $N=400$. 
Observe that the curves agree with a unique one! 
The same collapse is also detected for the caloric curve (not shown).
Let us recall the main motivation of this letter, namely 
the possible relation between statistical and temporal averages
for systems for which one cannot a priori define an extensive energy, such
the case we are considering now. In Fig. 2 we have also included 
the plot (solid line) of the theoretical predictions of the equilibrium 
values obtained, by means of the canonical ensemble formalism, 
in the HMF version ($\alpha = 0$ with $N$--scaled 
potential energy)\cite{Antoni}. 
What we now observe is that not only  different size and different $\alpha$
curves, with $0\le\alpha< 1$, collapse into a unique one, but they  
collapse precisely to the mathematically tractable extensive mean--field
limit! 
 A similar collapse had already been found for the Ising ferromagnet 
submitted to a Monte Carlo process\cite{Sergio} but this is,
as much as we know, the first time this effect can be confirmed in a conservative
model with deterministic dynamics.
Since slowly decaying long--range interacting systems are ubiquitous in nature,    
our results, if valid for all those systems such that $\alpha <d$, reveal a simple 
way of calculating quantities at the equilibrium.  

Next we briefly describe what happens for  $1 < \alpha < 2$.
There exists still an order--disorder transition but 
with a sensitive dependence on the value of $\alpha$, as can
be observed in Figs. 3a and 3b for the particular case $\alpha=1.5$.
Contrary to the previous case where the critical energy (above which  
the magnetization falls down to zero as $N$ increases) remains 
independent on $\alpha$ although slightly smaller than the theoretical
prediction,
here as the value of $\alpha$ approaches 2, the critical energy decreases 
until a finite value which can be non--zero\cite{Jump}.  

Finally, we analyze the region $\alpha > 2$. Here, as expected, 
the system behaves similarly as in the first--neighbor limit, i.e., 
it does never order since for any finite energy the magnetization goes 
down to zero as $N \to \infty$. 
In Fig. 4 we plot $T/\tilde{N}$  vs. $E/(N\tilde{N})$ obtained 
numerically (symbols) for $\alpha=2.5$ and $\alpha=5.0$, together with the 
theoretical values for both the limit $\alpha \to \infty$\cite{Escande} 
(solid line) and the HMF model\cite{Antoni} (dashed line). 
We see how the numerical results approach those of the $\alpha \to \infty$
limit.
  
Summarizing, we have studied the equilibrium behavior of a one-dimensional
conservative system of interacting particles as a function of the range
of the interactions $\alpha$. By integrating numerically the equations of
motion we have found three different classes of systems. 
For $0 \le \alpha <1$
the systems undergo a second order phase transition and the measured 
quantities (e.g., the critical energy)
when suitably scaled, {\em do not depend} on the value of $\alpha$.
For $1 < \alpha < 2$ the systems undergo also a second order 
phase transition but, unlike the previous case,
the critical energy and the
magnetization curve depend sensitively on the range of the interactions
$\alpha$. Finally, for $\alpha > 2$ the systems adopt the
first-neighbor behavior where there is no order
at finite temperature for $N\to\infty$.

It is also worth noting that an the $\tilde{N}$--scaling performed
over the results produced by Hamiltonian (\ref{hamiltonian})  
leads to the same results as those produced by 
an artificial ``extensive''  Hamiltonian  
constructed with $\tilde{N}$-scaled potential energies 
generalizing the HMF approximation.
For a correspondence between both treatments when looking at
dynamical aspects, time should be $\tilde{N}^{1/2}$--scaled.

Probably the non-trivial data collapse here reported for
$0 \le \alpha < 1$,
can be extended to $0 \le \alpha < d$ for an arbitrary dimension $d$, being 
(from the generalization of (\ref{Ntilde}) to $d$ dimesions)
$\tilde{N} \sim \frac{1}{1-\alpha/d} (N^{1-\alpha/d}-1)$. 
If true, our findings reveal a simple way of calculating macroscopic 
quantities at the equilibrium when long--range interactions are involved.
\\[5mm] \noindent
We acknowledge C. Tsallis for fruitful discussions.
This work was partially supported by FAPERJ, CNPq (Brazil),  
CONICET, CONICOR and SECYT-UNC (Argentina).

%
\section*{Captions for Figures}
{\bf Figure 1:}(a) Modulus of the magnetization, $|{\bf M}|$, 
as a function of the specific energy $E/N$;  
(b)  $|{\bf M}|$ as a function of the scaled
specific energy $E/(N\tilde{N})$;  
(c) scaled temperature $T/\tilde{N}$ as a function of $E/(N \tilde{N})$. 
The symbols correspond to numerical 
simulations in the microcanonical ensemble for $\alpha=0.5$ and different 
system sizes indicated on the figure.
Each symbol corresponds to an average of different 
initial conditions (typically 10). 
The solid lines correspond to the theoretical equilibrium predictions
for the HMF model (for which $\tilde N=1$, 
since the HMF Hamiltonian is already $N$-scaled). \\[5mm] \noindent
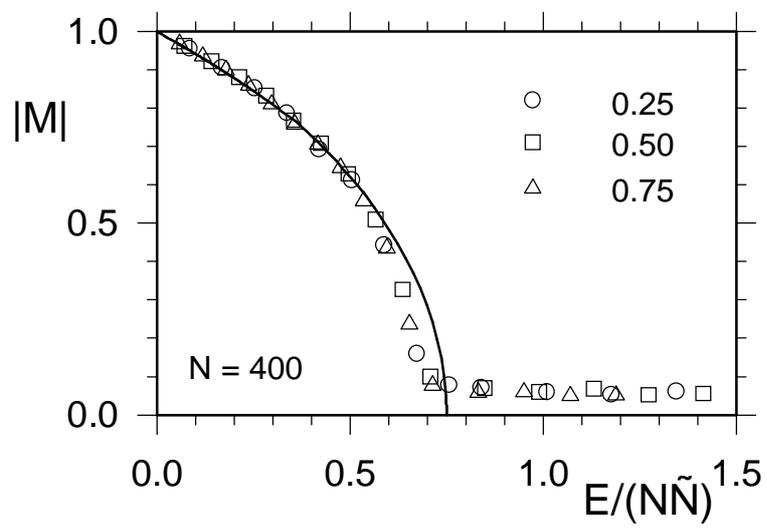
{\bf Figure 2:} $|{\bf M}|$ as a function of 
$E/(N \tilde{N})$. The symbols correspond to numerical 
simulations in the microcanonical ensemble  for $\alpha=0.25$, $0.5$, 
and $0.75$, all for $N=400$. The solid line corresponds to the theoretical 
results for the HMF model. \\[5mm] \noindent
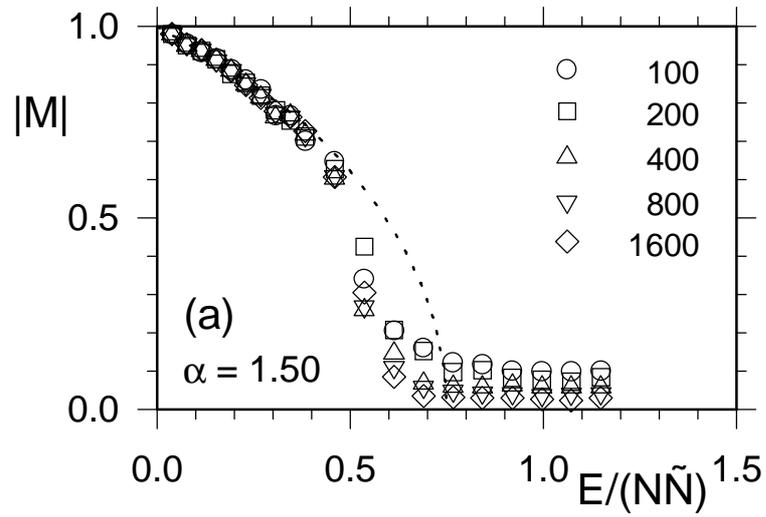
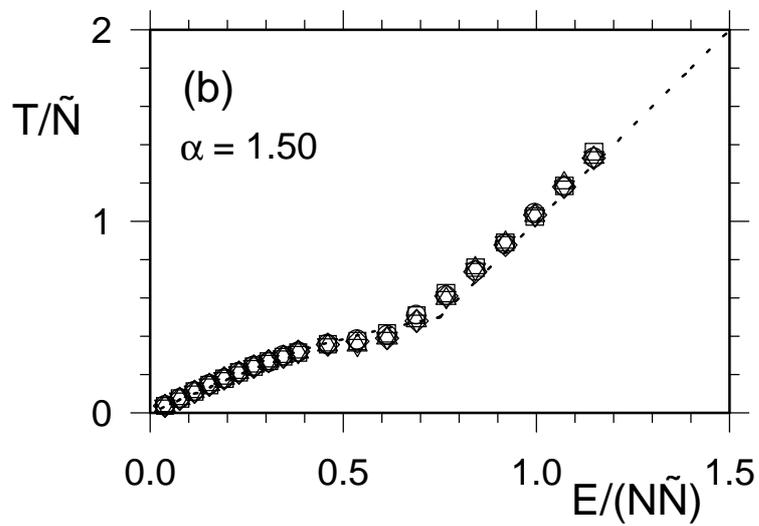
{\bf Figure 3:} 
(a)  $|{\bf M}|$ as a function of $E/(N \tilde{N})$;  
(b) scaled temperature $T/\tilde{N}$ as a function of $E/(N \tilde{N})$. 
The symbols correspond  to numerical simulations in the microcanonical ensemble 
for $\alpha=1.5$ and different system sizes.
For comparison, the dotted line corresponds to the theoretical 
equilibrium predictions of the HMF model. \\[5mm] \noindent
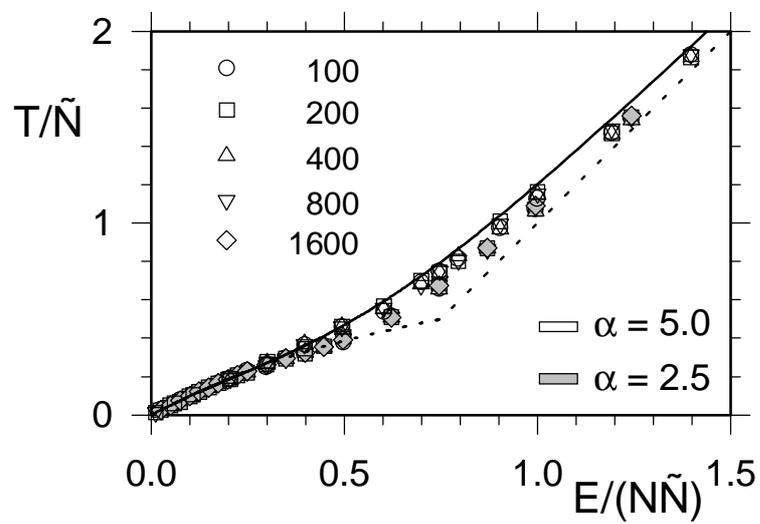
{\bf Figure 4:} 
Scaled temperature $T/\tilde{N}$ as a function of $E/(N \tilde{N})$. 
The symbols correspond to numerical simulations
in the microcanonical ensemble for $\alpha=$ $2.5$ (gray), $5.0$ (white) 
and different system sizes indicated on the figure.
The dashed and solid lines correspond to the theoretical results 
for the HMF ($\tilde{N}=1$) and the $\alpha\to \infty$ 
($\tilde{N}=2$) limits, respectively.

\end{multicols}


\begin{references}
%
\bibitem{Thirring} W. Thirring,  Foundations of Physics {\bf 20}, 1103 (1990).
%
\bibitem{Tsallis}  C. Tsallis,  Fractals {\bf 3}, 541 (1995).
\bibitem{Antoni} M. Antoni and S. Ruffo,  Phys. Rev. E {\bf 53},
 2361 (1995).
%
\bibitem{Vito1} V. Latora, A. Rapisarda and S. Ruffo, Phys. Rev. Lett.
{\bf 80}, 692 (1998).
%
\bibitem{Vito2} V. Latora, A. Rapisarda and S. Ruffo,
 Physica D {\bf 131}, 38 (1999); 
M. C. Firpo, Phys. Rev. E {\bf 57}, 6599 (1998).
%
\bibitem{Escande} D. Escande, H. Kantz, R. Livi and S. Ruffo, J. Stat.
Phys. {\bf 76}, 605 (1994).
%
\bibitem{Celia} C. Anteneodo and C. Tsallis,  Phys. Rev. Lett. 
{\bf 80}, 5313 (1998). 
%
\bibitem{Yoshida} H. Yoshida,  Phys. Lett. A {\bf 150}, 262 (1990).
%
\bibitem{algorithm} 
We employed an algorithm of order $N^2$. Notice, however, that the
algorithm could be improved using FFT since the Fourier base 
diagonalizes the Hamiltonian (Referee's comment). 
%
\bibitem{Sergio} S. A. Cannas and F. A. Tamarit,  Phys. Rev. B {\bf 54},
R12661 (1996).
%
\bibitem{Sampaio} L. C. Sampaio, M. P. de Albuquerque and F. S. Menezes, 
Phys. Rev. B {\bf 55}, 5611 (1997).
%
\bibitem{Jund} P. Jund, S. G. Kim and C. Tsallis, Phys. Rev. B {\bf 52}, 
50 (1995); J. R. Grigera, Phys. Lett. A {\bf 217}, 47 (1996); 
S. Curilef and C. Tsallis, preprint (1998).
%
\bibitem{Jump}  J. M. Kosterlitz, Phys. Rev. Lett. {\bf 37}, 1577 (1976);
S. A. Cannas, Phys. Rev. B {\bf 52}, 3034 (1995). 
%
\end{references}
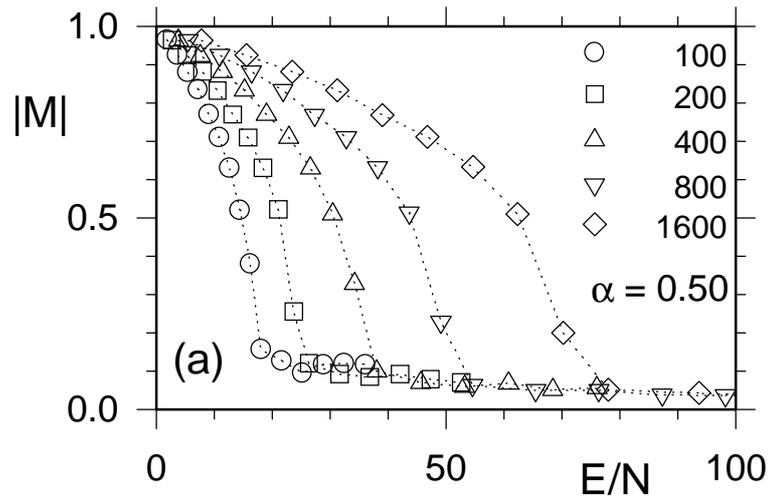
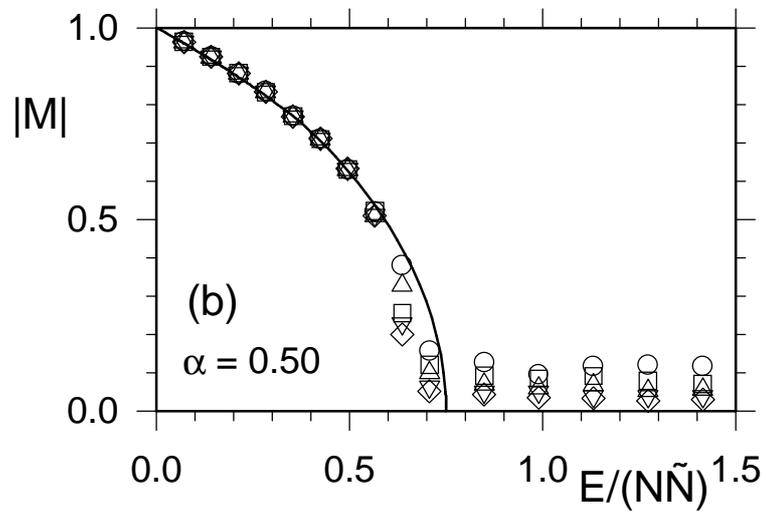
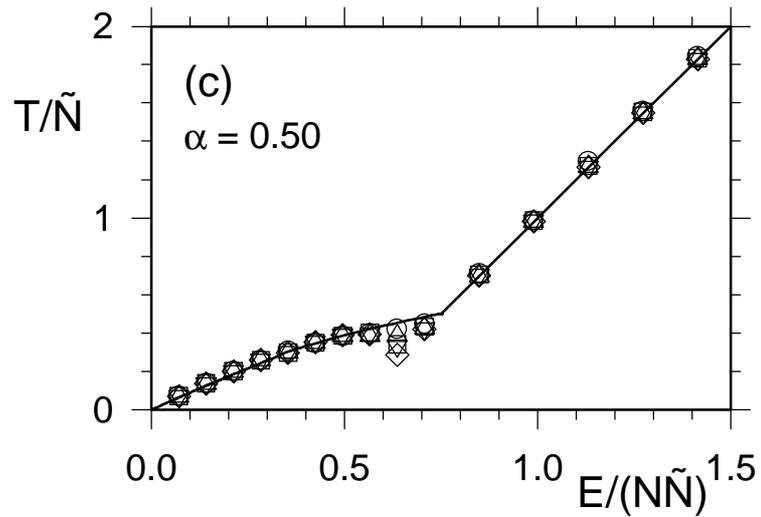
\end{document}